# Dynamical Properties of z~2 Star Forming Galaxies and a Universal Star Formation Relation[*]

Bouché[1], N., Cresci[1], G., Davies[1], R., Eisenhauer[1], F., Förster Schreiber[1], N.M., Genzel[1,2], R., Gillessen[1], S., Lehnert[1,9], M., Lutz[1], D., Nesvadba[1,9], N., Shapiro[3], K. L., Sternberg[4], A., Tacconi[1], L.J., Verma[5], A., Cimatti[6], A., Daddi[7], E., Renzini[8], A., Erb, D.K.[10], Shapley, A.[11] & Steidel, C.C. [12]

[1] Max-Planck-Institut für extraterrestrische Physik (MPE), Giessenbachstr.1, 85748 Garching, Germany
[2] Department of Physics, Le Conte Hall, University of California, 94720 Berkeley, USA
[3] Department of Astronomy, Campbell Hall, University of California, Berkeley, 94720, USA
[4] School of Physics and Astronomy, Tel Aviv University, Tel Aviv 69978, Israel
[5] Sub-Department of Astrophysics, University of Oxford, Denys Wilkinson Building, Keble Road, Oxford OX1 3RH, United Kingdom
[6] Universita di Bologna, Via Ranzani 1, I-40127 Bologna, Italy
[7] Laboratoire AIM, CEA/DSM - CNRS - Universite Paris Diderot, DAPNIA/SAp, 91191 Gif sur Yvette, France
[8] Osservatorio Astronomico di Padova, Vicolo dell'Osservatorio 5, Padova, I-35122, Italy
[9] GEPI, Observatoire de Paris, Place Jules Janssen, 92195 Meudon Cedex, France
[10] Harvard-Smithsonian Center for Astrophysics, 60 Garden Street, Cambridge, Mass. 02138, USA
[11] Princeton University Observatory, Peyton Hall, Ivy Lane, Princeton, NJ 08544, USA
[12] California Institute of Technology, MS 105-24, Pasadena, CA 91125, USA

## ABSTRACT

We present the first comparison of the dynamical properties of different samples of z~1.4-3.4 star forming galaxies from spatially resolved imaging spectroscopy from SINFONI/VLT integral field spectroscopy and IRAM CO millimeter interferometry. Our samples include 16 rest-frame UV-selected, 16 rest-frame optically-selected and 13 submillimeter galaxies (SMGs). We find that restframe UV- and optically bright (K<20) z~2 star forming galaxies are dynamically similar, and follow the same velocity-size relation as disk galaxies at z~0. In the theoretical framework of rotating disks forming from dissipative collapse in dark matter halos, the two samples require a spin parameter <λ> ranging from 0.06 to 0.2. In contrast bright SMGs ($S_{850\mu m} \geq 5$ mJy) have larger velocity widths and are much more compact. Hence, SMGs have lower angular momenta and higher matter densities than either of the UV- or optically selected populations. This indicates that dissipative major mergers may dominate the SMGs population, resulting in early spheroids, and that a significant fraction of the UV/optically bright galaxies have evolved less violently, either in a series of minor mergers, or in rapid dissipative collapse from the halo, given that either process may leads to the formation of early disks. These early disks may later evolve into spheroids via disk instabilities or mergers. Because of their small sizes and large densities, SMGs lie at the high surface density end of a universal (out to z=2.5) 'Schmidt-Kennicutt' relation between gas surface density and star formation rate surface density. The best fit relation suggests that the star formation rate per unit area scales as surface gas density to power ~1.7, and that the star formation efficiency increases by a factor 4 between non-starbursts and strong starbursts.

[*] based on observations at the Very Large Telescope (VLT) of the European Southern Observatory (ESO), Paranal, Chile, under programs GTO 073.B-9018, 074.A-9011, 075.A-0466, 076.A-0527, 077.A-0576, 078.A-0600, and 079.A-0341 and on observations obtained at the IRAM Plateau de Bure Interferometer (PdBI). IRAM is funded by the Centre National de la Recherche Scientifique (France), the Max-Planck Gesellschaft (Germany), and the Instituto Geografico Nacional (Spain).

## 1. Introduction

Deep surveys have become efficient in detecting z~1.5-3 star forming galaxy populations (e.g. Steidel et al. 1996, 2004, Franx et al. 2003, Daddi et al. 2004b, Chapman et al. 2005). Large samples are now available based on their rest-frame UV magnitude/color properties (the so-called 'BX/BM' criterion: Adelberger et al. 2004, Steidel et al. 2004, Erb et al. 2006a), or on their rest-frame optical magnitude/color properties ('star forming' or 's'-BzK: Daddi et al. 2004a,b, Kong et al. 2006; Gemini Deep Deep Survey (GDDS): Abraham et al. 2004), or on their submillimeter flux densities (Smail et al. 2002; Chapman et al. 2003,2005; Pope et al. 2006). The BX/BM or s-BzK selection criteria sample luminous (L~$10^{11-12}$ $L_\odot$) galaxies with star formation rates $\Re_*$~10-300 $M_\odot$/yr, and with ages from 50 Myrs to 2 Gyrs (Erb et al. 2006b, Daddi et al. 2004a,b). Bright submillimeter continuum selected systems (SMGs) with $S_{850\mu m}$>5mJy sample dusty, gas rich, and very luminous (~$10^{13}$ $L_\odot$) galaxies with much larger star formation rates $\Re_*$~$10^3$ $M_\odot yr^{-1}$ (Blain et al. 2002, Smail et al. 2002, Chapman et al. 2005).

It is obviously desirable to better understand the evolution of these different galaxy populations, as well as their relation to each other. A powerful new tool comes from spatially resolved dynamical studies. Such measurements are now feasible with near-IR integral field spectrographs (IFUs) on 8-meter class telescopes with SINFONI at the ESO/VLT (e.g. Forster Schreiber et al. 2006) and with OSIRIS on Keck (e.g. Wright et al. 2007), and with high resolution millimeter interferometry in CO rotation lines, for instance with the IRAM Plateau de Bure Interferometer (PdBI: Guilloteau et al. 1992).

In this paper, we present the first comparison of the different z~2 galaxy samples (UV, optical, submm) in terms of their structural and dynamical properties, combining the near-IR IFU and millimeter interferometry techniques. Throughout, we use a ΛCDM cosmology with $\Omega_M$=0.3, $\Omega_\Lambda$=0.7, Ho=70 km/s/Mpc.

## 2. Observations

As part of our Spectroscopic Imaging survey of high redshift galaxies in the Near-infrared with SINFONI (SINS) guaranteed time program, we have been studying the kinematic properties of a substantial sample of z~1.4-3.2 galaxies selected with the rest-frame UV- and optical selection techniques with integral field spectroscopy (Förster Schreiber et al. 2006, Genzel et al. 2006).

For the rest-frame UV sample, we observed seventeen galaxies drawn from the Hα survey of Erb et al. (2006a), which were selected by their UGR colors down to R<25.5. In the literature (Adelberger et al. 2004), these are referred to as `BX' galaxies (in the redshift range 2.0-2.43) and as `BM' galaxies (in the redshift range 1.0-1.5). We detect Hα for fifteen `BX' galaxies at <z>=2.24 and one BM galaxy at z=1.4. In the sole non-detection the Hα line happened to fall directly on a strong OH sky line, which prevented meaningful observations. Fourteen of these galaxies have already been presented in Förster Schreiber et al. (2006). As described in Förster Schreiber et al. (2006), the SINFONI BX/BM sample appears to be a fair representation of the larger Erb et al. (2006a) Hα sample in terms of source sizes and dynamical masses. Our selection criteria emphasize marginally brighter (<F(Hα)> ~$10^{-16}$ compared to $6 \times 10^{-17}$ erg/s/cm$^2$) and somewhat broader line width (<$v_c$>~175 ±68 km/s rms, compared to 140 km/s) systems than the average galaxy in the larger sample of Erb et al. (2006a).

We also observed nineteen optically-selected galaxies, sixteen of which are detected in Hα and presented for the first time in this paper. The sample consists of eleven $K_s$<20 galaxies meeting the s-BzK color criterion of Daddi et al. (2004b) [taken from the GOODS-S (Daddi et al. 2004a) and Deep 3a (Kong et al. 2006) fields] and of eight galaxies from the Gemini Deep Deep Survey (GDDS, Abraham et al. 2004) with $K_s$≤20.5 and I≤24.5. Thus, the nineteen optically selected galaxies have similar magnitude limits. From the flux-limited samples, actively star forming galaxies are selected with the 's'-BzK color criterion for the 11 BzK galaxies, and by requiring the presence of presence of the [OII] emission line, and young stellar populations ('100' category in Abraham et al.) for the GDDS galaxies. All s-BzK and five of the eight GDDS galaxies were detected in Hα. For the three GDDS galaxies, the non-detection of Hα is most likely due to a faint line emission and unfavorable wavelengths (close to bright sky OH emission line). Table 1 summarizes our sub-sample sizes.

For meaningful comparisons amongst the samples, we separate them in two redshift bins, z=2.0-2.4 and z=1.35-1.65. The K-selected and Hα detected sample comprises ten in the redshift range z=2.0 to 2.4, (<z>=2.23), and two from GDDS. Three BzK and three detected GDDS galaxies are at <z>=1.52 (see Table 1).

For all but three galaxies, we employed SINFONI (Eisenhauer et al. 2003) in seeing limited mode (0.125"x0.25" pixels), resulting in FWHM ~0.5" resolution (see Förster Schreiber et al. 2006). Three were observed with adaptive optics (as in Genzel et al. 2006) with 0.050"x0.100" pixels resulting in a FWHM of 0.15"-0.35". The spectral resolution is about 80 km/s and 100 km/s FWHM, in the K- and H-bands respectively. For a description of data reduction methods we refer to Schreiber et al. (2004), Abuter et al. (2006) and Förster Schreiber et al. (2006).

We compare our UV/optically selected galaxies with a sample of SMGs observed at the IRAM Plateau de Bure Interferometer (Downes & Solomon 2003, Genzel et al. 2003, Neri et al. 2003, Kneib et al. 2004, Greve et al. 2005, Tacconi et al. 2006). The SMG sample consists of ten bright ($S_{850\mu m} \geq 5 mJy$) SMGs in the z~2.2-3.4 range, from the radio selected sample of Chapman et al. (2003, 2005). We also include three SMGs from the work of Downes & Solomon (2003), Genzel et al. (2003) and Kneib et al. (2004). The median redshift of the fourteen SMGs is <z>=2.49.

**3. Results**

From the SINFONI data we extracted Hα fluxes, star formation rates, intrinsic half intensity radii ($R_{1/2}$), and circular velocities ($v_c$), which give dynamical masses within $R_{1/2}$. Following the same method as described in Förster Schreiber et al. (2006) we used both the source integrated velocity width, as well as the observed velocity gradient (detected in 14 of the 16 BX/BM, and 13 of the 16 s-BzK/GDDS galaxies), for deriving two independent estimates of $v_c$. To do so quantitatively we analyzed the data in the framework of simple axisymmetric, rotating disk models, taking into account beam smearing and inclination[1]. In the cases where inclination could not be determined from the Hα morphological aspect ratios, we adopted sin(i) =<sin(i)>=2/π.

The Hα detection rate and kinematic properties of the s-BzKs and GDDS galaxies appear slightly different. All eight z~2 s-BzKs are detected and well resolved in Hα, with large velocity gradients (>100 km/s). In contrast, only five (of eight) of the GDDS sources, which are mostly at lower redshifts, are detected in Hα. There is a tendency for all GDDS sources and two of the three, z~1.5 s-BzKs to be more compact, with small or modest velocity gradients (40-100 km/s). For the SMGs the PdBI observations have established a source size or a significant upper limit in $^{12}$CO line emission or 1mm continuum for eight of the galaxies (Tacconi et al. 2006). Only one of these shows a resolved velocity gradient in CO (SMMJ02399-0136, Genzel et al. 2003). The other observed SMGs are too compact for detecting kinematic structure at the presently best resolution of ~0.5". With these data in hand, we are now in the position to carry out a first comparison between the various samples in terms of their velocity and sizes (section 3.1), matter densities (section 3.2), and global star formation relations (section 3.3).

**3.1 Velocity-size relation: kinematic comparison of the samples**

Disk galaxies at z~0 exhibit a well known correlation in the velocity-size plane (Courteau

1997). The velocity-size plane is a powerful tool for constraining the angular momentum properties of our galaxies from a theoretical (e.g. Fall and Efstathiou 1980; Mo, Mao & White 1998, hereafter MMW) or observational point of view (e.g. Courteau 1997; Puech et al. 2007). In the framework of dissipative collapse to centrifugally supported disks of baryonic gas in dark matter halos (White & Rees 1978, MMW), the disk scale length $R_d$ is given by

$$R_d \propto v_c \lambda (j_d/m_d) H(z)^{-1} , \qquad (1)$$

where $v_c$ is the halo circular velocity, $\lambda$ is the spin parameter of the halo, $(j_d/m_d)$ the ratio of the disk angular momentum fraction, $j_d$, to the disk to halo mass fraction, $m_d$, and $H(z)$ is the Hubble parameter at redshift z. The proportionality constant depends on the specific dark matter halo profile and the assumed disk physics. Because $H(z)$ increases by a factor of 3 from z=2 to z=0, high-redshift disks ought to be smaller for a given $v_c$, if the disk spin parameter $\lambda'=\lambda (j_d/m_d)$ does not change with redshift. We assume that the baryons experienced no loss of angular momentum during collapse (Fall and Efstathiou 1980), which implies $j_d/m_d=1$, and $\lambda'=\lambda$. This condition is supported by our earlier results in Förster Schreiber et al. (2006).

In Figure 1 we show a comparison of the velocity-size properties between our different z~1.5-3.4 galaxies (left panel), and the z~0 Sbc disks from Courteau (1997) (right panel). In plotting the high redshift galaxies, we assume that the exponential disk scale length ($R_d$) of the Hα emission is equal to the inferred linear, intrinsic half light radius $R_{1/2}$. This assumption is supported by model disk fits incorporating the beam smearing due to the ~0.5" FWHM resolution of our data. The maximum circular velocity is determined from either the total velocity width and velocity gradients[1] (when available), or from disk modeling for five of our z~2 galaxies. Systematic uncertainties (included in the typical error given as a large cross in the right panel of Fig.1) are substantial, especially in the $R_d$ coordinate, which is strongly affected by the clumpy and often asymmetric Hα brightness distribution.

The immediate question is whether the z~2 galaxies truly are rotating disks in gravitational equilibrium. This question can presently be addressed quantitatively only for a subset of about a dozen or so of our BX/s-BzK galaxies for which good quality, two-dimensional kinematics can be extracted (Forster Schreiber et al. 2006, Genzel et al. 2006, Cresci et al. in prep.). Within this subset a rotating disk interpretation is compelling for more than half of the cases (Shapiro et al., in preparation) indicating that at least 50% of our sources are indeed disks. There are a few galaxies (at most four) where a merger interpretation is more appropriate, in which case the velocity-size plot is still useful in a virial sense.

Independent of the modeling details, there are two significant conclusions from Fig.1. First, the BX/BM and s-BzK/GDDS galaxies we have observed have comparable dynamical properties. The BX/BM and s-BzK criteria select largely overlapping populations (at least down to $K_s<20$), in excellent agreement with Reddy et al. (2005) and Grazian et al. (2007). Second, and in contrast to the rest-frame UV/optically selected galaxies, SMGs occupy a much different part of the $v_c$-$R_d$ plane. SMGs are more compact and at the same time have greater velocity widths (as in Swinbank et al. 2004) than the UV/optically selected galaxies (by a factor of ~2 in each coordinate: Greve et al. 2005, Tacconi et al. 2006, Iono et al. 2006). SMGs appear dynamically distinct from the UV/optically selected samples and appear to have low orbital angular momenta. Since gas rich, major mergers are efficient at removing angular momentum through large scale gravitational torques (Barnes & Hernquist 1996), it is tempting to conclude that this dichotomy is a result of whether or not a galaxy has recently undergone a major merger. Note that in a gas rich, major merger, the gas settles into a central disk rapidly, in $10^8$ yr (Barnes & Hernquist 1996, Iono, Yun, & Mihos 2004), justifying a rotating disk approximation used in Figure 1. In fact, even in early stage mergers, such as NGC 6240, the molecular gas is centered between the two radio/near-infrared nuclei (Tacconi et al. 1999, Iono et al. 2007) and is also the center of rotation and marks the maximum of velocity dispersion in both the stars and the gas.

---

[1] Models of rotating disks with appropriate parameters for BzK/BX/BM/GDDS galaxies and observed with ~0.5" FWHM suggest $v_c\sin(i)\sim0.42 \Delta v$(FWHM) and $v_c\sin(i)\sim0.67 v_r$, where $\Delta v$ is the total FWHM line width of the galaxy and $v_r$ is half of the maximum velocity gradient observed on either side of the galaxy (Förster Schreiber et al. 2006).

Surprisingly, the right panel of Fig.1 shows that the rest-frame UV/optically selected z~2 galaxies also overlap with the velocity-size distribution of the z~0 disks of Courteau (1997), as also found from recent IFU results on z≤0.6 star forming galaxies (Puech et al. 2007). The spread in the $v_c$-$R_d$ plane is comparable to that of z~0 disks. The five z~2 galaxies with full 2 dimensional modeling confirm that the overlap with the z~0 galaxies is not an artifact of our data. The overlap with the z=0 sample suggests that there exists a significant population of z~2 galaxies with total angular momenta comparable to z~0 disk galaxies. In contrast, our z~2 galaxies do not overlap with the distribution of local early type galaxies in this plane (e.g. in a comparison with the data from Bender et al. 1992).

The solid line in the left panel of Fig. 1 represents the best fit halo angular momentum parameter λ (=0.08) to our UV/optically selected galaxies, in the limit of isothermal ($v_c$=const) dark matter halo profiles in eq. (1). The dotted and long-dashed curves in Fig. 1 represent models obtained from eq. (1) with more realistic NFW (Navarro, Frenk and White 1997) dark matter halos, with the inclusion of self-gravity of the disks as prescribed in MMW, and with the z-scaling of halo concentration parameters found by Bullock et al. (2001): $c_{vir}$=9 $(M_{vir}/1.1 \times 10^{13} M_\odot)^{-0.13}/(1+z)$.

Such 'MMW' models with λ=0.06 and λ=0.21 bracket our high-z galaxies. The average λ-parameter we deduce is twice the average angular momentum parameter found in dark matter simulations (λ=0.04–0.05; e.g. MMW). Observations of a larger sample are required to test whether this result is an artifact of our selections, or truly an intrinsic property of the high-z star forming galaxies. In addition, to be so actively star forming the baryonic disks must be strongly self-gravitating, and may even be unstable under their own gravity, with $m_d$≥λ (MMW). This implies that the galaxies we are observing have disk mass fractions $m_d$ of ~0.08 on average, corresponding to about half of the cosmic baryon fraction ($\Omega_{baryon}/\Omega_{dark\ matter}$~0.18: Spergel et al. 2006).

### 3.1.1 Additional considerations

In comparing SMGs with UV-/optically selected galaxies, it is important to establish whether Hα on the one hand, and CO/mm continuum emission on the other hand are good tracers of the global distribution of gas and star formation in dusty star forming galaxies. Since SMGs appear to be 'scaled-up' (more luminous and more gas rich) versions of z~0 ultra-luminous infrared galaxy (ULIRG) (Tacconi et al. 2006), one can gain insights on the SMG properties from the z~0 analogs. For the ULIRG population, the answer to the above question is unequivocally positive in terms of the CO/mm-continuum emission (c.f. Tacconi & Lutz 2001). It is thus very plausible that the mm-line and continuum emission trace the distribution of star formation in SMGs as well. Local ULIRGs have small effective radii (inferred from near-IR photometry, tracing the mass), a few kpc or less (Genzel et al. 2001, Tacconi et al. 2002), which are comparable to the CO sizes (Downes & Solomon 1998, Tacconi et al. 1999, Iono et al. 2007). In SMGs CO/mm-continuum sizes are also comparable to those in the radio continuum (Tacconi et al. 2006, Chapman et al. 2004), which is likely another good tracer of star formation via the FIR-radio correlation. Thus, the sizes inferred from CO/mm-continuum for SMGs are likely representative of the size of the host. In a given SMG, CO/mm-continuum emission and Hα emission do not need to originate from the same region(s).

SMGs/ULIRGs have the largest dust column densities and extinctions and thus are the worst cases for Hα observations. UV-/optically selected galaxies likely have much lower dust content and extinction than SMGs and it seems thus safer to assume that Hα is a good tracer of the overall star formation distribution in these sources.

Similarly, it is important to establish, for the SMGs, whether the line widths measured from the CO line emission are representative of the gravitational potential, and whether (AGN driven) outflows play any significant role. Double-peaked profiles occur when two close nuclei are orbiting each other, in a rotating nuclear ring, or disk. Such profiles have been observed in z~0 ULIRGs (Narayanan et al. 2005), such as Arp 220, and the majority (50%) of SMGs (Tacconi et al. 2006). In the case when rotation dominates, the two profiles are broad and symmetric. Narayanan et al. (2006) showed that the kinematic signature of AGN-driven outflows are asymmetric profiles both in terms of line width of each component and relative intensity. However, the outflow signature does not dominate the overall kinematics, and may be visible only in 25% of the time (Narayanan et al. 2006). Given that in local ULIRGs, signature of molecular outflows have not been observed, and that the AGN contribution to the SMGs FIR flux is small (see section 3.3), it is extremely unlikely that the molecular line emission in SMGs is affected by molecular outflows.

In their study of a larger sample of $S_{850\mu m} \geq 5$ mJy SMGs with radio detections (from which our SMG sample was drawn to a large extent), Chapman et al. (2005) find that the majority of SMGs with spectroscopic redshifts have restframe UV colors compatible with the 'BX' criterion of Steidel et al. (2004). This overlap in color is compatible with the dichotomy we find from our dynamical measurements for the following reason. The average R-band magnitude of the SMGs in the Chapman et al. (2005) sample is 24.9, or about 1.3 mag fainter than the average R-band magnitude of our BX sources. The 'BX' (or s-BzK)-component of an SMG comes from low extinction star forming regions that contribute only a small fraction of the average star formation rate (less than 10%) similarly to local ULIRGs (Tacconi & Lutz 2001). Our dynamical dichotomy between submillimeter and rest-frame optically-UV selected sources thus is not at odds with the optical/UV photometric properties of the two populations.

### 3.2 SMGs are much denser than UV/optically selected galaxies

Figure 2 shows a comparison of the various high-z samples and MMW disk models in the dynamical disk mass-volume density plane. As in Figure 1, UV- and optically selected galaxies and SMGs are clearly separated, while star forming BX/BM and s-BzK/GDDS galaxies overlap. To the extent that CO and Hα measurements provide fair and comparable estimates of the dynamical properties of each galaxy, the separation between z~2 UV/optically selected galaxies and SMGs is almost all due to their matter density and not due to the mass enclosed within the half light radius of the CO/Hα data, except for galaxies in the z~1.5 sub-sample, which have smaller dynamical masses. Volume/surface densities of UV/optically selected galaxies are similar to those in z~0 disks. SMGs are on average one to two orders of magnitude denser, their average matter densities range from 100 to 1000 cm$^{-3}$. As we will show in the next section, they also have high surface densities (see also Nesvadba et al. 2007). Such high surface/matter densities are higher than local spheroid densities (Bender et al. 1992) and similar to that of z~1.5 spheroids, which are also smaller and denser than local spheroids (Trujillo et al. 2006, Longhetti et al. 2007). The circular velocities (200-500 km/s) and potential wells of SMGs are similar to z~0 spheroids as already found by Tacconi et al. (2006) and by Swinbank et al. (2006).

### 3.3 A universal 'Schmidt-Kennicutt' relation

In Figure 3, we show the star formation rate surface density [$\Sigma_{starform}=\Re_*/(\pi R_{1/2}^2)$] as a function of gas surface density [$\Sigma_{gas}$] (left panel) and of gas density divided by dynamical time $\tau_{dyn}=R_{1/2}/v_c$ (right panel). The star formation rates $\Re_*$ for the UV/optically selected galaxies are from our H$\alpha$ luminosities, corrected for extinction (Erb et al. 2006a,b, Daddi et al. 2004b) and applying $\Re_*=L(H\alpha)/7.4\times10^{40}$ erg/s. The latter relationship is taken from Kennicutt (1998; hereafter K98) with a correction factor of 0.6 to convert the 0.1-100 $M_\odot$ Salpeter initial mass function (IMF) used by K98 to a Chabrier IMF. Star formation rates for the SMGs are from 850μm flux densities and $\Re_*=110\ S_{850}$ (mJy), derived from the Pope et al. (2006) average conversion from $S_{850}$ to FIR luminosity and a continuous star formation model. SMGs with $S_{850}\geq5$ mJy form stars at a rate an order of magnitude greater than the UV/optically selected galaxies, given that recent Spitzer IRS spectra of SMGs have shown that the SMGs are indeed powerful starbursts and not Xray-obscured AGNs (Lutz et al, Valiante et al. 2007, Pope et al. 2007). The gas surface densities ($\Sigma_{gas}$) are computed using measured gas fractions ($f_{gas}$) from the molecular gas content for the SMGs, and assuming $f_{gas}=0.4$ for the rest-frame UV and optically selected galaxies, as described below.

Figure 3 shows that there is a fairly tight relationship between $\Sigma_{starform}$ and $\Sigma_{gas}$ (left), or $\Sigma_{gas}/\tau_{dyn}$ (right). Combining our high-z galaxy data with the local star forming galaxies (grey crosses) in K98, we find that there appears to be a universal Schmidt-Kennicutt relationship, independent of redshift out to z=2.5. The larger star formation rates of SMGs can be understood almost solely as a consequence of their smaller sizes and larger surface densities compared to those of UV/optically selected galaxies.

In constructing Figure 3 we made an important assumption that differs from K98, which changes the resulting correlations quantitatively but not qualitatively. A number of studies during the last decade have yielded compelling evidence that the conversion factor from $^{12}$CO line intensity to total molecular hydrogen gas mass drops significantly, and perhaps suddenly, beyond gas surface densities of $\sim10^2$ $M_\odot pc^{-2}$ (Wild et al. 1992, Solomon et al. 1997, Downes & Solomon 1998, Oka et al. 1998, Rosolowsky et al. 2003, Davies, Tacconi & Genzel 2004, Hinz & Rieke 2006). As a consequence we took, as K98, a Galactic conversion factor [$X_G=2-3\times10^{20}$ cm$^{-2}$/(K km/s)] for the normal galaxies in the K98 sample but adopted $X=0.25X_G$ for the starbursts/(U)LIRGs in K98, which is an average of the values obtained in the above references for high gas density/star formation density galaxies. For the high-z galaxies we derived gas surface densities from the dynamical mass densities ($\Sigma_{gas}=f_{gas} M_{dyn}(R\leq R_{1/2})$ / ($\pi R_{1/2}^2$), with $M_{dyn}=v_c^2R_{1/2}/G$) and assumed a constant gas fraction of 0.4, which is motivated by the gas fraction found in SMGs with the same $X=0.25X_G$ (U)LIRG conversion factor (Greve et al. 2005, Tacconi et al. 2006). With a Galactic conversion factor, the gas masses of local ULIRGs and SMGs would exceed their dynamical masses by a factor of 2. The above relations for the 'X'-factor are appropriate for solar metallicity. Theoretical considerations (Maloney & Black 1988, Wall 2006) suggest that the conversion factor from CO flux to gas mass may increase with decreasing metallicity. However, there is no unanimous conclusion from the existing literature whether (Wilson 1995, Arimoto et al. 1996) or not (Rosolowsky et al. 2003) such an effect is seen in the Local Group of galaxies and how large it is. We have, therefore, assumed the solar metallicity conversion factor throughout.

For the Chabrier (2003) stellar mass function used here, the best fit power-laws to both low- and high-z galaxies yield universal Schmidt-Kennicutt relations of the forms

$\Sigma_{starform}$ ($M_\odot yr^{-1}kpc^{-2}$) = 9.3 (±2) x $10^{-5}$ ($\Sigma_{gas}$ ($M_\odot pc^{-2}$) )$^{1.71(\pm0.05)}$ and,

$\Sigma_{starform}$ ($M_\odot yr^{-1}kpc^{-2}$) = 0.037 (±0.004) ($\Sigma_{gas}/\tau_{dyn}$ ($M_\odot yr^{-1}kpc^{-2}$) )$^{1.14(\pm0.03)}$     (2),

where error bars are 1σ uncertainties to the fit. The power law index in the first relation is significantly greater than given in K98 ($\Sigma_{starform}\sim\Sigma_{gas}^{1.4\pm0.15}$). It is also marginally greater than the free-fall index of 1.5. This is perhaps suggestive of a collisional or pressure enhancement of the star formation rate at high densities. In contrast to equation 7 of K98 ($\Sigma_{starform} \sim 0.02\ \Sigma_{gas}/\tau_{dyn}$) the second relationship implies a density dependent star formation efficiency. Equation (2) implies that z~2 SMGs and z~0 ULIRGs have a ~4 times greater star formation efficiency than normal galaxies like the Milky Way.

The agreement between low- and high-z Kennicutt-Schmidt laws is remarkable given our assumption of a constant gas fraction and extinction corrections based on UV/optically data alone. Erb et al. (2006b) have in fact plausibly explained some of the outliers in Fig.3 as these are found to have a very small (≤0.1) ratio of stellar mass to dynamical mass and thus likely gas fractions higher than assumed here.

**4 Discussion**

From our dynamical comparison (Figure 1), it appears that rest-frame UV (BX/BM) and optically selected star forming galaxies (s-BzK) have similar properties, and probably are drawn from overlapping populations down to Ks<20, in agreement with the conclusions reached by Reddy et al. (2005) and Grazian et al. (2006) on the basis of optical/UV photometry.

The most important differences in dynamical/star forming properties between UV/optical star forming galaxies and SMGs are the smaller sizes (Figure 1), and much larger surface/volume densities of the SMGs (Figure 2). This strong dichotomy may be due to whether or not a galaxy has recently undergone a major merger, resulting in a large loss of angular momentum, and whether the resulting remnant has been transformed into a dense, spheroidal component. This scenario is supported by the comparison of the SMG properties to those of spheroids and by recent high resolution mm-interferometry, which shows that several of the SMGs in Figs.1 could be major mergers (Tacconi et al. 2006, and in preparation), as suspected for some time from statistical arguments (Smail et al. 2002) and morphological studies (Conselice, Chapman & Windhorst 2003).

Given that the kinematics of UV and optical star forming galaxies are consistent with that of large protodisks, these systems may have evolved in a less violent fashion through a series of minor mergers, or through cold gas flows from the halo (Kereš et al. 2005, Dekel & Birnboim 2006). In the simplest evolutionary model of dissipative accretion, disk scale lengths would be expected to increase by a factor $1/H(z)$, or 3 between z=2 and z=0 in a ΛCDM cosmology. Taking into account the much smaller concentration parameters at high-z suggested by the Bullock et al. (2001) scaling this growth factor would be smaller but would still push the larger, more massive z~2 disks way above the location of most z~0 disks (right panel of Fig.1). A still smaller growth factor would be predicted if the baryonic disk mass fraction increases with decreasing redshift (Somerville et al. 2006). Alternatively and perhaps more likely, some fraction of the z~2 disks we are observing may have been subsequently destroyed and converted to spheroids by mergers or secular evolution (van den Bosch 2002, Immeli et al. 2004, Förster Schreiber et al. 2006, Genzel et al. 2006, Governato et al. 2006), in agreement with the cosmic volume density of BX/s-BzK and their clustering (Adelberger et al. 2004, 2005, Kong et al. 2006). Substantial further growth of the disks may also have been inhibited by large negative feedback from AGN/star formation (van den Bosch 2002, Burkert & Lin 2006), or by shock quenching when the halos exceeded a mass of a few $10^{12}\ M_{\odot}$ (Dekel & Birnboim 2006, Birboim et al. 2007). Future dynamical studies with higher spatial resolution will explore the redshift range z~1.5 further to test whether the possible trend in the UV/optically selected galaxies for smaller sizes and circular velocities at z~1.5 (relative to ~2.3) is a consequence of this evolutionary scenario.

Given that SMGs with $S_{850} \geq 5$ mJy have small sizes and large matter densities, they lie at the high surface density end of a universal (out to z=2.5) 'Schmidt-Kennicutt' relation (Figure 3). Since the Schmidt-Kennicutt relation implies that the star formation rate per gas particle increases with gas surface density, SMGs must naturally be more luminous than their UV-/optically selected kin and form stars at a rate an order of magnitude greater than the UV/optically selected galaxies despite similar dynamical masses.

*Acknowledgements. We are grateful to the staff of Paranal Observatory for their support, and the SINFONI team for their hard work on the instrument that has made this research possible. We also thank the 'SMG team', A.Blain, F.Bertoldi, S.Chapman, P.Cox, T.Greve, R.Ivison, R.Neri, A.Omont and I.Smail. We acknowledge useful discussions with A.Burkert, A.Dekel, T.Naab and E.Quataert. We thank the anonymous referee for his/her critical comments that led to an improved manuscript.*

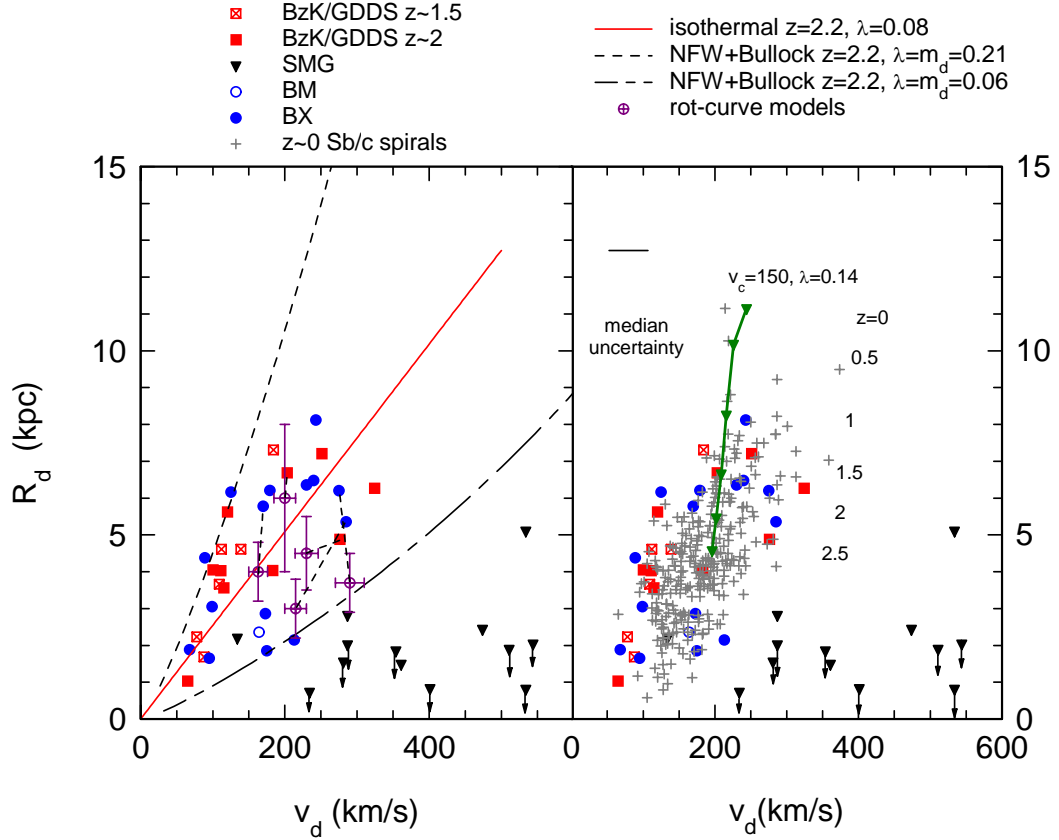

Fig.1. **Left:** For the different z~2 galaxy samples, we show the exponential disk scale length ($R_d$) vs. the maximum rotation velocity ($V_d$) (filled circles (BX), open circles (BM), filled triangles (SMGs), filled squares (z~2 BzK/GDDS) and open crossed squares (z~1.5 BzK/GDDS)). Whenever possible, an inclination correction was estimated from the intrinsic morphological aspect ratio or assuming $<\sin i>=2/\pi$, as appropriate for a statistical average. For five z~2 galaxies we also carried out a full disk modeling and here we plot the best fit disk scale length and inclination corrected maximum disk circular velocity (crossed open circles). Heavy dotted lines connect these better estimates with the simpler ones above. The continuous line is the best fit to the BX and BzK/GDDS data of an isothermal model of a centrifugally supported disk within a halo of spin parameter of $\lambda=0.08$. The dash-dotted and dotted curves are MMW models for self-gravitating disks in NFW halos for $\lambda=0.06$ and $0.21$ where the disk mass fraction $m_d=\lambda$ (critical disks). **Right:** Comparison of the high-z galaxy sample with the sample of z~0 Sb/c disk galaxies of Courteau (1997, grey crosses). The solid line with triangular markers denotes the redshift evolution of a disk from z=2.5 to z=0 for a $v_c=150$ km/s NFW halo with $\lambda'=m_d=0.14$ (with concentration parameter as in the text). The median uncertainty of the high-z data is indicated at the upper left of the right panel, including fit errors, and also systematic uncertainties in inclination correction and the transformation from observed FWHM sizes to disk scale lengths. Size upper limits are marked as arrows for the SMGs.

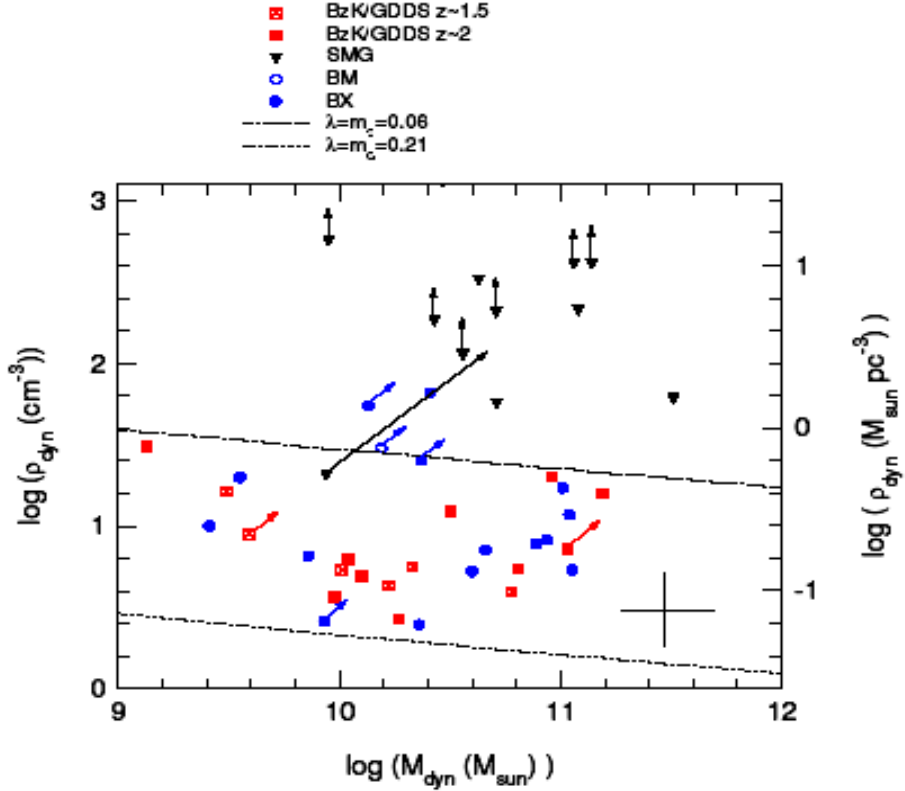

Fig 2. Global matter volume density ($\rho_{dyn}=M_{dyn}/(2\pi(R_{1/2})^2 h_z)$), with z-scale height $h_z$, as a function of dynamical mass within $R_{1/2}$, for the various samples. The dashed and dash-dotted curves are the two bracketing MMW models from Figure 1, and symbols are the same as in Figure 1. We assumed a ratio of $h_z/R_{1/2} \sim 0.25$, as inferred by Tacconi et al. (2006), Förster Schreiber et al. (2006), Genzel et al. (2006) and Elmegreen & Elmegreen (2006). Size upper limits are marked as vertical arrows. Inclination limits are marked as diagonal arrows. The SMG SMM14011+0252 is likely at low inclination and its true location may be significantly to the right (long arrow). The SMGs have matter densities (>100 cm$^{-3}$) similar to that of local ellipticals and spheroids (Kormendy 1989).

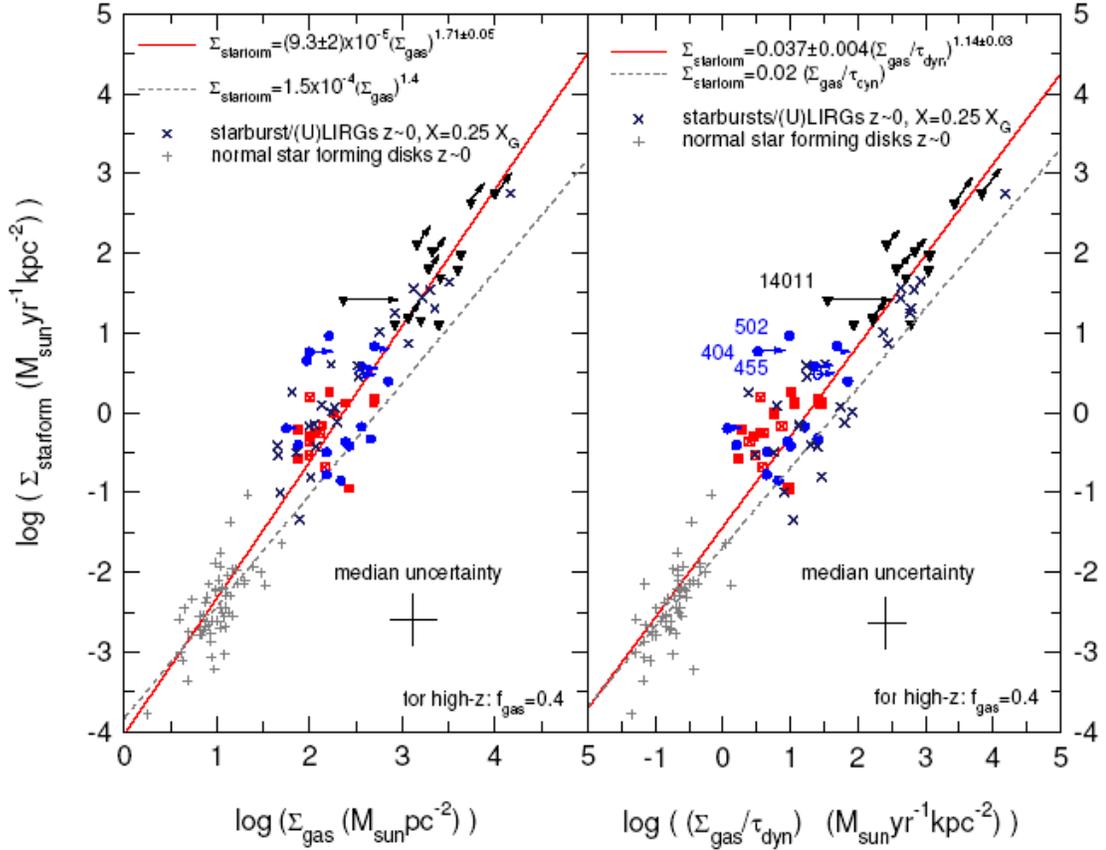

Fig.3. Schmidt-Kennicutt relations for low- and high-z galaxies. The symbols for high-z galaxies are the same as in Figure 1. Grey crosses and stars denote normal galaxies and starbursts/luminous infrared galaxies from Kennicutt (1998) (K98, see their Figs. 6 & 7). The median uncertainty of the data points (including uncertainties in inclination correction) is denoted by a large cross. The vertical axis is the star formation rate surface density, $\Sigma_{starform}=\Re_*/(\pi R_{1/2}^2)$, for a Chabrier (2003) stellar mass function. The horizontal axis in the left panel is the gas surface density, and in the right panel the gas surface density divided by dynamical time $\tau_{dyn}=R_{1/2}/v_c$. The continuous red curve is the best-fit power law to all low- and high-z data excepting the outlier SMM14011+0252 (Nesvadba et al. 2007). The dotted grey curves are the relationships proposed in Kennicutt (1998: $\Sigma_{starform}\sim 0.02\ \Sigma_{gas}/\tau_{dyn}$ and $\Sigma_{starform}=1.5\times 10^{-4}(\Sigma_{gas})^{1.4}$), which assume a constant CO-$H_2$ conversion factor for all galaxies. Of the three prominent BX outliers above the correlation two (BX1623-455, BX1623-502, denoted in the Figure) are found by Erb et al. (2006a) to have a very small ($\leq 0.1$) ratio of stellar mass to dynamical mass and thus, very likely gas fractions higher than assumed here. BX404 and the SMG SMM14011+0252 are likely at low inclination and their true location may be significantly to the right.

**Table 1. Summary of our subsample sizes. The detections refer to Hα for the rest-frame UV and optically selected galaxies and to CO for the SMGs.**

| Samples | Observed | Detected | z=2-2.5 | z=1.35-1.65 |
|---|---|---|---|---|
| BX/BM | 17 | 16 | 15 | 1 |
| K20 (GOODS) | 5 | 5 | 5 | 0 |
| s-BzK (Deep3a) | 6 | 6 | 3 | 3 |
| GDDS | 8 | 5 | 2 | 3 |
| SMG | 13 | 13 | 13 | 0 |
| Total | 49 | 45 | 38 | 7 |